\documentclass[pra,preprint,superscriptaddress,amsmath,amssymb,showpacs]{revtex4}
\usepackage{bm}
\usepackage{amsmath}
\usepackage{color}
\usepackage[dvips]{graphicx}

\begin{document}

\title{Angular distribution of high-energy $e^+e^-$ photoproduction close to the end of spectrum}

\author{A. Di Piazza}
\email{dipiazza@mpi-hd.mpg.de}
\affiliation{Max-Planck-Institut f\"ur Kernphysik, Saupfercheckweg 1, 69117 Heidelberg, Germany}

\author{A. I. Milstein}
\email{milstein@inp.nsk.su}
\affiliation{Max-Planck-Institut f\"ur Kernphysik, Saupfercheckweg 1, 69117 Heidelberg, Germany}
\affiliation{Budker Institute of Nuclear Physics of SB RAS, 630090 Novosibirsk, Russia}

\date{\today}

\begin{abstract}
We consider the differential cross section of electron-positron pair production by a high-energy photon in a strong Coulomb field close to the end of the electron or positron spectrum. When the momentum transfer largely exceeds the electron mass, the cross section is obtained analytically in a compact form. Coulomb corrections essentially modify the cross section even for moderate values of the nuclear charge number $Z$. In the same kinematical region, the angular distribution for bound-free pair production, bremsstrahlung, and photorecombination is also obtained.
\end{abstract}

\maketitle

\section{Introduction}
The process of electron-positron ($e^+e^-$) pair production by a photon in a strong atomic field has been investigated since many years (see the reviews \cite{HGO1980,Hubbell2000}). The cross section of this process in the leading order in $Z\alpha$ (Born approximation), is known for arbitrary energy $\omega$ of the incoming photon \cite{BH1934,Racah1934}. Here, $Z$ is the atomic charge number and $\alpha\approx 1/137$ is the fine-structure constant (units with $\hbar=c=1$ are employed throughout). The formal expression of the cross section of $e^+e^-$ pair photoproduction, exact in the parameters $\eta=Z\alpha$ and $\omega$, was derived in \cite{Overbo1968}. This expression has a very complicated form which leads to substantial difficulties in numerical computations. The difficulties grow as $\omega$ increases, so that numerical results have been obtained so far only for $\omega<12.5\,$ MeV \cite{SudSharma2006}.
In the high-energy region $\omega\gg m$, with $m$ being the electron mass, a simple expression of the cross section was obtained in \cite{BM1954,DBM1954}, exactly with respect to $\eta$ and in the leading approximation in $m/\omega$. However, this expression provides rather accurate results only at energies $\omega \gtrsim 100\,$MeV. On the other hand, the theoretical description of the total cross section at $\eta\lesssim 1$ and at intermediate photon energies between $5\,$ MeV and $100\,$ MeV has been based for a long time on the extrapolation of the results obtained at $\omega<5\,$MeV, Ref.\cite{Overbo1977}. Finally, results for the spectrum of one of the created particles at intermediate photon energies were practically absent. An important step has been made recently in \cite{LMS2004} in this direction, where the first corrections of the order of $m/\omega$ to the spectrum as well as to the total cross section of $e^+e^-$ photoproduction in a strong atomic field were derived. The correction to the spectrum was obtained in the region where both produced particles are ultrarelativistic. In \cite{AA2010}, the spectrum was obtained in the region where one of the produced particles is ultrarelativistic and the other has an energy of the order of the electron mass. Essentially less is known on the angular distribution of the final particles at intermediate photon energies. Coulomb corrections, i.e., the contributions of higher-order terms of the perturbation theory with respect to $\eta$, are much more important for the angular distribution than for the spectrum. In \cite{BM1954,DBM1954}, the angular distribution of $e^+e^-$ photoproduction was obtained exactly in the parameter $\eta$ in the leading order with respect to $m/\omega$. This result was obtained under the assumption that both created particles are ultrarelativistic and that the angles between their momenta and the momentum of the initial photon are small. Under the same assumptions, the first quasiclassical correction to the angular distribution found in \cite{BM1954,DBM1954}, was derived very recently in \cite{LMS2011}. In this paper a noticeable charge asymmetry in the differential cross section of high-energy $e^+e^-$ photoproduction was predicted. The angular distribution, when one of the particles is not ultrarelativistic, is not known for $\eta\lesssim 1$ although one can expect Coulomb corrections to be important in this case. The investigation of this problem for arbitrary angles between the momenta of the final particles and the photon momentum is in general a complicated task. In the present paper, we consider a particular case of this problem, which admits a relatively simple analytical solution. Namely, for high-energy $e^+e^-$ photoproduction in a strong Coulomb field, we investigate the distribution over the angle $\theta$ between the positron momentum $\bm p_+$ and the photon momentum $\bm k$ at electron energies $\epsilon_-$ much smaller than the positron energy $\epsilon_+$, so that the electron may not be ultrarelativistic. We also assume that
\begin{equation}
\label{condition}
\frac{\epsilon_-}{\epsilon_+}\ll\theta\ll\sqrt{\frac{\epsilon_-}{\epsilon_+}},
\end{equation}
which implies a large transverse positron momentum $Q=\omega\theta\approx\epsilon_+\theta\gg \epsilon_-$. Finally, in the same kinematical region we have also obtained the angular distribution for bound-free pair production, where the final electron is in an arbitrary bound state, as well as for the cross channels, i.e., for bremsstrahlung and for photorecombination.

\section{Calculation of the cross section}
In order to calculate the cross section of $e^+e^-$ photoproduction, differential over the angles of the fast positron and integrated over the angles of the slow electron, we can employ the relation
\begin{equation}
\label{compl}
\sum_{\lambda}\int \frac{d\Omega}{4\pi}\Psi_{\lambda}(\bm{p},\bm{r})\Psi^{\dag}_{\lambda}(\bm{p},\bm{r})=\frac{\pi}{p^2}\sum_{j,\sigma,\mu}U_{j,\sigma,\mu}(p,\bm r)U^{\dag}_{j,\sigma,\mu}(p,\bm r)
\end{equation}
between the positive-energy electron states $\Psi_{\lambda}(\bm{p},\bm{r})$ in a Coulomb field with definite momentum $\bm{p}$ and polarization index $\lambda$, and those $U_{j,\sigma,\mu}(p,\bm r)$ with definite total angular momentum $j$, projection $\mu$ on some quantization axis, and parity $-\sigma$ (see, e.g., \cite{MS1983,AA2010}). In Eq. (\ref{compl}) $d\Omega$ is the solid angle corresponding to the direction of the momentum $\bm{p}$. In this way, the cross section, averaged over the polarization of the incoming photon and summed up over the polarizations of the final electron and positron, has the form (see, e. g., Ref. \cite{AA2010})
\begin{eqnarray}\label{sigmaee0}
\frac{d\sigma}{d\bm p_+}&=&-\frac{\alpha}{8\pi^2}\frac{1}{\omega\beta_-}\sum_{\lambda_+,\,j,\,\sigma,\,\mu}{\cal M}^{\rho}{\cal M}^*_{\rho}
\,,\nonumber\\
{\cal M}^{\rho}&=&
\int d\bm r\,\bar U_{j,\sigma,\mu}(p_-,\bm r)\gamma^\rho V^{(+)}_{\lambda_+}(\bm p_+,\bm r)\mbox{e}^{i\bm k\cdot\bm r}\,.
\end{eqnarray}
In these equations $(\omega,\bm k)$ is the four-momentum of the photon, $\beta_-=p_-/\epsilon_-$ is the modulus of the electron velocity and $\gamma^\rho$ are the Dirac matrices. The wave function $V^{(+)}_{\lambda_+}(\bm p_+,\bm{r})$ is the negative-energy wave function in a strong Coulomb field with $\bm p_{+}$ and $\lambda_+$ being the positron momentum and its polarization index. The asymptotics of this wave functions at large $r$ contains a plane wave and a divergent spherical wave \cite{BLP1982}.

For the negative-energy wave function $V^{(+)}_{\lambda_+}(\bm p_+,\bm r)$ corresponding to an ultrarelativistic particle, one can use the Furry-Sommerfeld-Mauer form \cite{BLP1982} 
\begin{equation}\label{FSMV}
V^{(+)}_{\lambda_+}(\bm p_+,\bm r) = \mbox{e}^{\pi\eta/2}\Gamma(1-i\eta)\mbox{e}^{-i\bm p_+\cdot\bm r}\left(1+\frac{i}{2\epsilon_+}\bm{\alpha}\cdot\bm{\nabla}\right)F(-i\eta,1,i(p_+r+\bm p_+\cdot\bm r))v_{\lambda_+},
\end{equation}
where $\bm{\alpha}=\gamma^0\bm{\gamma}$, $F(a,b,z)$ is the confluent hypergeometric function and where
\begin{equation}
\label{v}
v_{\lambda_+}=\begin{pmatrix}
\bm\sigma\cdot\bm \zeta\,\chi_{\lambda_+}\\
\chi_{\lambda_+}
\end{pmatrix}\,,
\end{equation}
with $\bm \zeta=\bm p_+/p_+$, $\bm{\sigma}$ being the Pauli matrices, and $\chi_{\lambda_+}$ being a constant spinor. For the sake of convenience, we choose the system of reference such that the vector $\bm{p}_+$ points along the positive $z$ axis. Thus, $(\bm{k}-\bm{p}_+)\cdot \bm{r}\approx-r(Q\sin\vartheta\,\cos\varphi+\epsilon_-\cos\vartheta)$, where $Q=\omega\theta$ is the modulus of the momentum transfer, $\vartheta$ is the angle between the vectors $\bm p_+$ and $-\bm r$, and $\varphi$ is the azimuth angle of $-\bm{r}$ in the plane perpendicular to $\bm{p}_+$. The main contribution to the quantity ${\cal M}^\rho$ in Eq. ({\ref{sigmaee0}) is given by the region of integration $r\sim 1/\epsilon_-$ and $|\sin\vartheta|\sim \epsilon_-/Q\ll 1$, so that either $\pi-\vartheta\sim \epsilon_-/Q$ or $\vartheta\sim \epsilon_-/Q$. However, for $\pi-\vartheta\sim \epsilon_-/Q$ the argument of the hypergeometric function is very large (of the order of $\epsilon_+/\epsilon_-$), which makes the integrand highly oscillating for such values of $\vartheta$. Therefore, the largest contribution to ${\cal M}^\rho$ comes from the region $\vartheta\sim \epsilon_-/Q\ll 1$. In this region the argument of the confluent hypergeometric function is of the order of $\omega\vartheta^2/\epsilon_-\sim \epsilon_-/(\omega\theta^2)$, which is much larger than unity due to our condition in Eq. (\ref{condition}), but much smaller than $\epsilon_+/\epsilon_-$. As a result, we can use the asymptotics of $V^{(+)}_{\lambda_+}(\bm p_+,\bm r)$ in Eq. (\ref{FSMV}), which is nothing but the eikonal form of this wave function,
\begin{equation}\label{wfV}
V^{(+)}_{\lambda_+}(\bm p_+,\bm r) = (p_+r+\bm p_+\cdot\bm r)^{i\eta} \mbox{e}^{-i\bm p_+\cdot\bm r}v_{\lambda_+}.
\end{equation}

The wave function $U_{j,\sigma,\mu}(p_-,\bm r)$ has the following form \cite{BLP1982}:
\begin{eqnarray}\label{wfU}
&&U_{j,\sigma,\mu}(p_-,\bm r)=
\begin{pmatrix}
f(r)\Omega_{j,l,\mu}(\bm n)\\
-\sigma \,g(r) \Omega_{j,l',\mu}(\bm n)
\end{pmatrix}\, ,\nonumber\\
&&f(r)=\frac{\sqrt{2}}{r}\sqrt{1+\frac{m}{\epsilon_-}}\,\mbox{e}^{(\pi\nu/2)}\frac{|\Gamma(\gamma+1+i\nu)|}{\Gamma(2\gamma+1)}
(2p_-r)^\gamma\,\mbox{Im} \left\{e^{i(p_-r+\xi)}F(\gamma-i\nu,2\gamma+1,-2ip_-r)\right\}\,,\nonumber\\
&&g(r)=\frac{\sqrt{2}}{r}\sqrt{1-\frac{m}{\epsilon_-}}\,\mbox{e}^{(\pi\nu/2)}\frac{|\Gamma(\gamma+1+i\nu)|}{\Gamma(2\gamma+1)}
(2p_-r)^\gamma\,\mbox{Re} \left\{e^{i(p_-r+\xi)}F(\gamma-i\nu,2\gamma+1,-2ip_-r)\right\}\,,\nonumber\\
&&l=j+\frac{\sigma}{2}\,,\quad l'=j-\frac{\sigma}{2}\,,\quad \nu=\dfrac{\eta}{\beta_-}\,,\quad \kappa=\sigma \left(j+\frac{1}{2}\right)\,,\quad
\gamma=\sqrt{\kappa^2-\eta^2}\,,\nonumber\\
&&\xi=(1-\sigma)\frac{\pi}{2}+\arctan\left[\frac{\nu(\epsilon_--m)}{\epsilon_- (\gamma+\kappa)}\right]\,, \quad\sigma=\pm 1\,,\quad
\mbox{e}^{-2i\xi}=\frac{\kappa+i\nu m/\epsilon_-}{\gamma+i\nu}\,,\, \bm n=\frac{\bm r}{r}
\end{eqnarray}
where $\Omega_{j,l,\mu}(\bm n)$ is a spherical spinor. We direct the quantization axes for the electron spin along the vector $\bm \zeta$. We recall that each component of the spherical spinors $\Omega_{j,j\pm 1/2,\mu}(\vartheta,\varphi)$ is proportional either to $\sin^{\mu-1/2}\vartheta$ or to $\sin^{\mu+1/2}\vartheta$ \cite{BLP1982}. Thus, since $\vartheta\ll 1$, the main contribution to the sum over $\mu$ comes from the terms with $\mu=\pm 1/2$, and the wave function $U_{j,\sigma,\mu}(p_-,\bm r)$ can be written as
\begin{eqnarray}\label{wfU1}
&&U_{j,\sigma,1/2}(p_-,\bm r)=\sqrt{\frac{j+1/2}{4\pi}}(-i)^{j-\sigma/2}\,
\begin{pmatrix}
if(r)\varphi_{1/2}\\
- \,g(r)\varphi_{1/2} 
\end{pmatrix}\, ,\nonumber\\
&&U_{j,\sigma,-1/2}(p_-,\bm r)=\sigma\,\sqrt{\frac{j+1/2}{4\pi}}(-i)^{j-\sigma/2}\,
\begin{pmatrix}
-if(r)\varphi_{-1/2}\\
- \,g(r)\varphi_{-1/2} 
\end{pmatrix}\, ,\nonumber\\
&&\varphi_{1/2}=\begin{pmatrix}
1\\
0
\end{pmatrix}\, ,\quad \varphi_{-1/2}=\begin{pmatrix}
0\\
1
\end{pmatrix}\, .
\end{eqnarray}
It is convenient to introduce the functions $F$ and $G$ as
\begin{eqnarray}\label{FG}
&&\begin{pmatrix}
F\\
G
\end{pmatrix}\, =\sqrt{\frac{j+1/2}{4\pi}}\int d\bm r\,(p_+r+\bm p_+\cdot\bm r)^{i\eta} \mbox{e}^{i(\bm k-\bm p_+)\cdot\bm r}\,
\begin{pmatrix}
f(r)\\
g(r)
\end{pmatrix}\, .
\end{eqnarray}
It can be easily shown that in terms of these functions, the cross section  (\ref{sigmaee0}) has the simple form
\begin{eqnarray}\label{sigmaee1}
\frac{d\sigma}{d\bm p_+}&=&\frac{\alpha}{4\pi^2}\frac{1}{\omega\beta_-}\sum_{j,\,\sigma}\left[|F|^2+|G|^2+2\mbox{Im}(FG^*)\right]
\,.
\end{eqnarray}
Under the condition (\ref{condition}), we can make the replacement in Eq. (\ref{FG}) (see the discussion below Eq. (\ref{v})),
\begin{equation}
p_+r+\bm p_+\cdot\bm r\longrightarrow \frac{1}{2}\omega r\vartheta^2\,,\quad (\bm k-\bm p_+)\cdot\bm r\longrightarrow 
-r(Q\vartheta\cos\varphi+\epsilon_-).
\end{equation}
Then we take the integral over $\varphi$, $\vartheta$ and $r$:
\begin{eqnarray}\label{FGphithetar}
&&\begin{pmatrix}
F\\
G
\end{pmatrix}\, =-\frac{4\pi i\eta}{Q^2}\sqrt{\frac{j+1/2}{4\pi}}\left(\frac{2\omega}{Q^2}\right)^{i\eta}\,\frac{\Gamma(1+i\eta)}{\Gamma(1-i\eta)}
\int_0^\infty dr\, r^{-i\eta}\mbox{e}^{-i\epsilon_-r}\,
\begin{pmatrix}
f(r)\\
g(r)
\end{pmatrix}\nonumber\\
&&=-\frac{2^{\gamma+1}\pi i\eta}{Q^2}\sqrt{\frac{j+1/2}{2\pi}}\left(\frac{2\omega p_-}{Q^2}\right)^{i\eta}\,
\frac{|\Gamma(\gamma+1+i\nu)|\Gamma(\gamma-i\eta)\Gamma(1+i\eta)}{\Gamma(2\gamma+1)\Gamma(1-i\eta)}\left(\frac{\beta_-}{1-\beta_-}\right)^{\gamma-i\eta}\,\nonumber\\
&&\times\mbox{e}^{\pi(\nu-\eta-i\gamma)/2}\,
\begin{pmatrix}
-i\sqrt{1+\frac{m}{\epsilon_-}}(\mbox{e}^{i\xi}{\cal F}_1-\mbox{e}^{-i\xi}{\cal F}_2)\\
\sqrt{1-\frac{m}{\epsilon_-}}(\mbox{e}^{i\xi}{\cal F}_1+\mbox{e}^{-i\xi}{\cal F}_2)
\end{pmatrix}\,,
\end{eqnarray}
where
\begin{equation}
 {\cal F}_1=F(\gamma-i\eta,\, \gamma-i\nu,\,2\gamma+1,\,-x)\,,\quad
{\cal F}_2=F(\gamma-i\eta,\, \gamma+1-i\nu,\,2\gamma+1,\,-x)\,,\quad x=\frac{2\beta_-}{1-\beta_-}\,,
\end{equation}
with $F(a,b,c,z)$ being the hypergeometric function. By substituting this result in Eq. (\ref{sigmaee1}) and by performing the summation over $\sigma$, we finally obtain,
\begin{eqnarray}\label{sigmaeefinal}
&&\frac{d\sigma}{d\bm p_+}=\frac{2\alpha}{\pi}\frac{\eta^2}{\omega\beta_-Q^4}\mbox{e}^{\pi(\nu-\eta)}
\sum_{j}\left(j+\frac{1}{2}\right)\frac{|\Gamma(\gamma+1+i\nu)|^2 |\Gamma(\gamma-i\eta)|^2}{\Gamma^2(2\gamma+1)}x^{2\gamma}\nonumber\\
&&\times\left[(1-\beta_-)|{\cal F}_1|^2+(1+\beta_-)|{\cal F}_2|^2
+2\nu (1-\beta_-^2)\mbox{Im}\left(\frac{{\cal F}_1^*{\cal F}_2}{\gamma+i\nu}\right)\right]
\,.
\end{eqnarray}
The result for the analogous cross section $d\sigma/d\bm p_-$ of photoproduction at $\epsilon_-\gg\epsilon_+$ is given by Eq. (\ref{sigmaeefinal}) with the replacements $\eta\rightarrow -\eta$, $\beta_-\rightarrow\beta_+$, and $\nu=\eta/\beta_-\rightarrow -\eta/\beta_+$.

\section{Discussion of the results}
At $\epsilon_+\gg \epsilon_-$, $\nu\ll 1$, and $Q=\omega\theta\gg \epsilon_-$, the cross section integrated over the angles of the electron momentum $\bm p_-$ can be easily found from the general expression of the cross section in the Born approximation (see, e.g., \cite{BLP1982}). It has the form,
\begin{eqnarray}\label{sigmaeefinalas2}
&&\frac{d\sigma_B}{d\bm p_+}=\frac{2\alpha}{\pi}\frac{\eta^2}{\omega Q^4}\,\ln\left(\frac{1+\beta_-}{1-\beta_-}\right)\,.
\end{eqnarray}
For $\eta\ll 1$ our result in Eq. (\ref{sigmaeefinal}) is in agreement with this formula. Although the above expression of $d\sigma_B/d\bm p_+$ tends to zero at $\beta_-\to 0$, the cross section (\ref{sigmaeefinal}) in the limit $\beta_-\to 0$ at fixed $\eta$ (when $\nu\to\infty$), is not zero. The most convenient way to obtain this last asymptotics is to substitute the asymptotics 
\begin{eqnarray}\label{fgbeta0}
&&f(r)=\frac{2\sigma\sqrt{2\pi\eta\beta_- }}{u}\left[(\kappa-\gamma)J_{2\gamma}(2\sqrt{u})+\sqrt{u}J_{2\gamma+1}(2\sqrt{u})\right]\,,\nonumber\\
&&g(r)=\frac{2\sigma\eta\sqrt{2\pi\eta\beta_-} }{u}J_{2\gamma}(2\sqrt{u})\,,\quad u=2\eta mr\,.
\end{eqnarray}
of the functions $f(r)$ and $g(r)$ at $\beta_-\to 0$ in Eq. (\ref{FGphithetar}) (see Appendix in Ref. \cite{AA2010} and note the different definition of the functions $f(r)$ and $g(r)$ there). As a result we obtain
\begin{eqnarray}\label{FGas1}
&&\begin{pmatrix}
F\\
G
\end{pmatrix}\, =-\frac{2\pi i\sigma\sqrt{\beta_-(j+1/2)}}{Q^2}\left(\frac{2\omega m}{Q^2}\right)^{i\eta}\,
\frac{\Gamma(\gamma-i\nu)\Gamma(1+i\eta)}{\Gamma(1-i\eta)\Gamma(2\gamma+1)}\,(2\eta)^{\gamma+1/2}\nonumber\\
&&\times\mbox{e}^{-\pi(\eta+i\gamma)/2}\,
\begin{pmatrix}
(\kappa-\gamma){\cal G}_1-2i(\gamma-i\eta){\cal G}_2\\
\eta{\cal G}_1
\end{pmatrix}\,,\nonumber\\
&&  {\cal G}_1=F(\gamma-i\eta,\,2\gamma+1,\,2i\eta)\,,\quad
{\cal G}_2=\frac{\eta}{2\gamma+1}F(\gamma+1-i\eta,\,2\gamma+2,\,2i\eta)\,.
\end{eqnarray}

By employing these formulas, we arrive at the following asymptotics of the cross section at $\beta_-\to 0$ and fixed $\eta$,
\begin{equation}\label{sigmaeefinalas1}
\frac{d\sigma}{d\bm p_+}=\frac{4\alpha\mbox{e}^{-\pi\eta}}{\omega Q^4}
\sum_{j}\left(j+\frac{1}{2}\right)^3 (2\eta)^{2\gamma+1} \frac{|\Gamma(\gamma-i\eta)|^2}{\Gamma^2(2\gamma+1)}
\left[|{\cal G}_1|^2+2|{\cal G}_2|^2
-2\mbox{Im}\left({\cal G}_1^*{\cal G}_2\right)\right]
\,.
\end{equation}
This expression shows that $d\sigma/d\bm p_+$ has a finite limit at $\beta_-\to 0$. At $\eta\ll 1$,  Eq. (\ref{sigmaeefinalas1}) becomes
\begin{eqnarray}\label{sigmaeefinalas11}
&&\frac{d\sigma}{d\bm p_+}=\frac{8\alpha\eta^3}{\omega Q^4}\,.
\end{eqnarray}
The ratio of $d\sigma/d\bm p_+$ at $\beta_-\to 0$, Eq. (\ref{sigmaeefinalas1}) and of its small-$\eta$ limit, Eq. (\ref{sigmaeefinalas11}) is displayed in Fig. \ref{beta0} as a function of $\eta$.
\begin{figure}
\begin{center}
\includegraphics[width=0.5\linewidth]{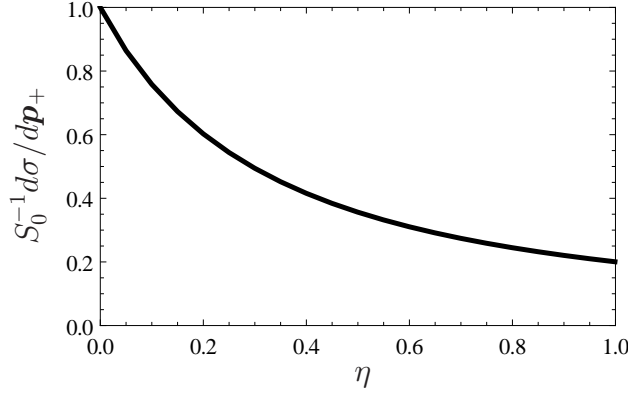}
\caption{The cross section $d\sigma/d\bm p_+$ at $\beta_-\to 0$ (see Eq. (\ref{sigmaeefinalas1})) in units of $S_0=8\alpha\eta^3/\omega Q^4$ as a function of $\eta$.}\label{beta0}
\end{center}
\end{figure}
It is seen that the contribution of high-order terms in $\eta$ essentially modifies the result obtained in the lowest order in $\eta$. Note that the asymptotics of the cross section $d\sigma/d\bm p_-$ at $\beta_+\to 0$ cannot be obtained from the asymptotic value in Eq. (\ref{sigmaeefinalas1}) via the replacement $\eta\rightarrow -\eta$. In fact, the result, following from the general formula in Eq. (\ref{sigmaeefinal}) after the replacements $\eta\rightarrow -\eta$, $\beta_-\rightarrow\beta_+$, and $\nu=\eta/\beta_-\rightarrow -\eta/\beta_+$, vanishes in the limit $\beta_+\to 0$ because the positron wave functions are exponentially small in this limit.

In part a) of Fig. \ref{beta1} the difference $d\sigma/d\bm p_+-d\sigma_B/d\bm p_+$ as a function of $\beta_-$ is plotted in units of $S_1=\alpha\eta^2/(\omega  Q^4)$ in the region $\beta_-$ close to unity, but $\epsilon_+$ still much larger than $\epsilon_-$. The difference  $d\sigma/d\bm p_--d\sigma_B/d\bm p_-$, with $d\sigma_B/d\bm p_-$ given by Eq. (\ref{sigmaeefinalas2}) with the replacement $\beta_-\rightarrow\beta_+$, is shown in part b) of Fig. \ref{beta1} as a function of $\beta_+$ for $\beta_+$ close to unity. The figure shows that in both cases the Coulomb corrections tend to zero as $\beta_{\pm}\to 1$. This fact can be explained as follows. The main contribution to the Coulomb corrections is given by the region of integration over distances $r$ of the order of the Compton wavelength $\lambda_C=1/m$, but in our kinematics both $r$ and $\rho=r\vartheta\sim 1/Q$ are much smaller than $\lambda_C$ at $\beta_-\to 1$ but still $\epsilon_+\gg\epsilon_-$, or at $\beta_+\to 1$ but still $\epsilon_-\gg\epsilon_+$, because in both cases $r\sim 1/\min(\epsilon_-,\epsilon_+)$. Also, as expected, the Coulomb corrections tend to increase the cross section with respect to the Born value in the case of fast positron (part a)) and to decrease it in the case of fast electron (part b)). It is interesting to note that at $\eta$ of the order of unity the Coulomb corrections are not symmetric even at $\beta_{\pm}$ close to unity.
\begin{figure}
\begin{center}
\includegraphics[width=\linewidth]{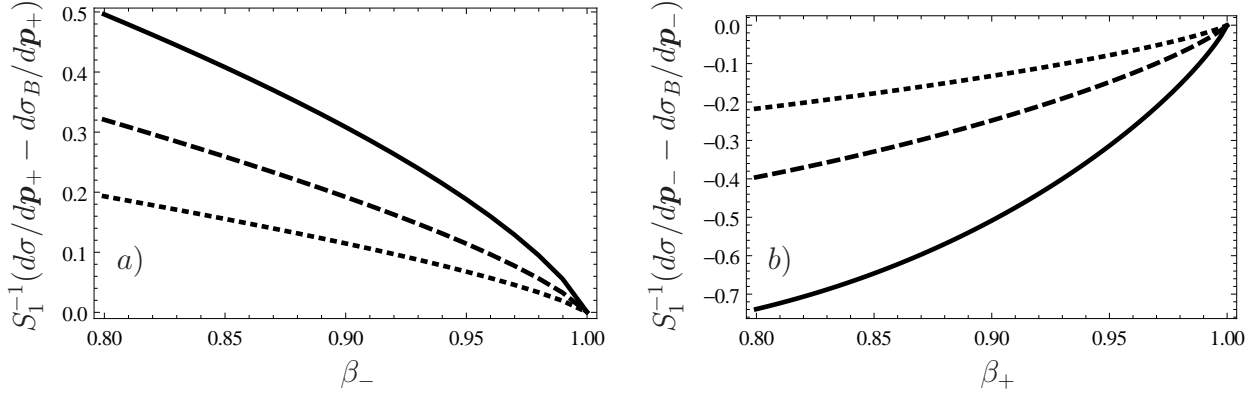}
\caption{The differences $d\sigma/d\bm p_+-d\sigma_B/d\bm p_+$ as a function of $\beta_-$ for $\beta_-$ close to unity (part a)) and  $d\sigma/d\bm p_--d\sigma_B/d\bm p_-$ as a function of $\beta_+$ for $\beta_+$ close to unity (part b)) plotted in units of $S_1=\alpha\eta^2/(\omega  Q^4)$. In each part the solid curve corresponds to $Z=92$, the dashed curve to $Z=47$ and the dotted curve to $Z=26$.}\label{beta1}
\end{center}
\end{figure}
Finally, in Fig. \ref{betaall} the cross section $d\sigma/d\bm p_+$ as a function of $\beta_-$ (part a)) and the cross section $d\sigma/d\bm p_-$ as a function of $\beta_+$ (part b)) are shown in units of $S_1$ in the whole interval of values of $\beta_-$ and $\beta_+$, and for a few values of the charge number $Z$. In both cases higher-order terms in $\eta$ play an important role in the whole interval of $\beta_{\pm}$ except that in a narrow region close to the point $\beta_{\pm}=1$.
\begin{figure}
\begin{center}
\includegraphics[width=\linewidth]{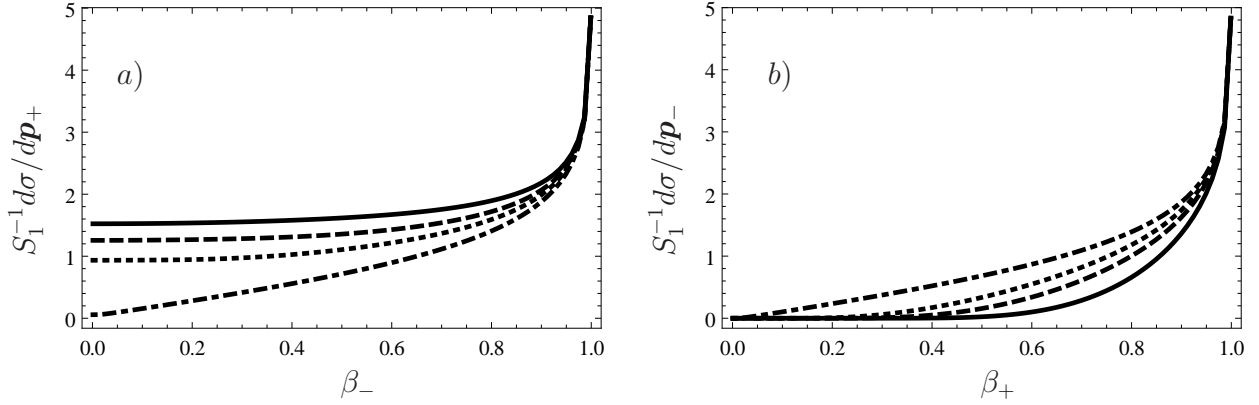}
\caption{The cross section $d\sigma/d\bm p_+$ as a function of $\beta_-$ (part a)) and  the cross section $d\sigma/d\bm p_-$ as a function of $\beta_+$ (part b)). In each part the solid curve corresponds to $Z=92$, the dashed curve to $Z=47$, the dotted curve to $Z=26$ and the dashed-dotted curve to $Z=1$.}\label{betaall}
\end{center}
\end{figure}

\section{Bound-free pair photoproduction and cross channels}

We consider now the high-energy photoproduction cross section with the electron in a bound state having total angular momentum $j$, projection $\mu$ on some quantization axis,  parity $-\sigma$, and radial quantum number $n_r$ (see \cite{MS93,AS97}, the review \cite{BS2006} and the references therein). The cross section $d\sigma_{bf}/d\Omega_+$ for $d\Omega_+$ being the solid angle corresponding to the positron momentum $\bm{p}_+$, averaged over the polarization of the incoming photon and summed up over the polarizations of positron and over $\mu$, has the form,
\begin{eqnarray}\label{sigmaeebf}
\frac{d\sigma_{bf}}{d\Omega_+}&=&-\frac{\alpha\omega}{4\pi}\sum_{\lambda_+,\,\mu}{\cal N}^{\rho}{\cal N}^*_{\rho}
\,,\nonumber\\
{\cal N}^{\rho}&=&
\int d\bm r\,\bar U_{j,\sigma,\mu,\,n_r}(\bm r)\gamma^\rho V^{(+)}_{\lambda_+}(\bm p_+,\bm r)\mbox{e}^{i\bm k\cdot\bm r}\,,
\end{eqnarray}
where $V^{(+)}_{\lambda_+}(\bm p_+,\bm{r})$ is given by Eq. (\ref{wfV}) and $U_{j,\sigma,\mu,\,n_r}(\bm r)$ is the positive-energy wave function of the bound state \cite{BLP1982}. Performing the same calculation as above, we obtain at $Q\gg m$,

\begin{equation}\label{sigmaeebffinal}
\begin{split}
\frac{d\sigma_{bf}}{d\Omega_+}=&\frac{2\alpha\eta^2 m\omega(j+1/2)}{Q^4}\left(\frac{2\eta}{N}\right)^{2\gamma+1}
\frac{\Gamma(2\gamma+n_r+1) |\Gamma(\gamma-i\eta)|^2}{\Gamma^2(2\gamma+1)n_r!}\\
&\times\exp\left[-2\eta \arctan\left(\frac{\gamma+n_r}{\eta}\right)\right]\left\{|{\cal F}_{b1}|^2+\frac{n_r}{2\gamma+n_r}|{\cal F}_{b2}|^2
+2n_r\mbox{Im}\left[\frac{{\cal F}_{b1}^*{\cal F}_{b2}}{\eta+i(\gamma+n_r)}\right]\right\}
\,.
\end{split}
\end{equation}
Here we have introduced the notation
\begin{eqnarray}\label{sigmaeebffinalnotation}
&&{\cal F}_{b1}=F(-n_r,\,\gamma-i\eta,\,2\gamma+1,\,y)\,,\quad {\cal F}_{b2}=F(1-n_r,\,\gamma-i\eta,\,2\gamma+1,\,y)\,,\nonumber\\
&&y=\frac{2\eta}{\eta+i(\gamma+n_r)}\,,\quad N=\sqrt{n_r^2+2\gamma n_r+\kappa^2}\,.
\end{eqnarray}
The cross section $d\sigma_{bf}/d\Omega_+$ in units of the quantity $S_{bf}=\alpha m\omega\eta^5/Q^4$ and summed up over $j$ and $n_r$, is shown as a function of $\eta$ in Fig. \ref{figbf} (solid line). The dashed line in this figure indicates the contribution of the ground state ($n_r=0$ and $j=1/2$) to the total cross section. The leading-order value of $d\sigma_{bf}/d\Omega_+$ at $\eta\ll 1$ in the above units gives $8\zeta(3)\approx 9.616$.
\begin{figure}
\begin{center}
\includegraphics[width=0.5\linewidth]{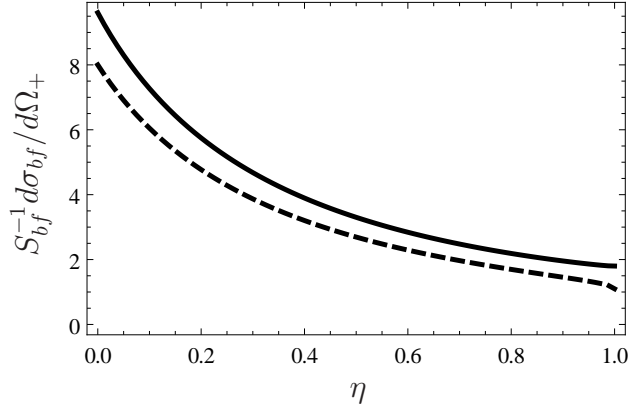}
\caption{The cross section  $d\sigma_{bf}/d\Omega_+$ of the bound-free pair photoproduction in units of $S_{bf}=\alpha m\omega\eta^5/Q^4$ as a function of $\eta$. The solid curve corresponds to the sum over all $j$ and $n_r$ and the dashed one to the contribution of the ground state ($n_r=0$ and $j=1/2$).}\label{figbf}
\end{center}
\end{figure}
Fig. \ref{figbf} shows that the Coulomb corrections essentially modify the cross section already at moderate values of $\eta$.

We conclude by briefly discussing the cross sections in the cross channels of $e^+e^-$ photoproduction. The cross section $d\sigma/d\bm k$ of bremsstrahlung of an ultrarelativistic initial electron with energy $\epsilon_1$ in a strong Coulomb field, when the final electron energy $\epsilon_2$ is much smaller than $\epsilon_1$ and $Q\gg\epsilon_2$, is given by the right-hand-side of Eq. (\ref{sigmaeefinal}), with the replacement $\omega\rightarrow\epsilon_1$ and
$\beta_-\rightarrow\beta_2=p_2/\epsilon_2$. Here, we assumed the result to be averaged over the polarization of the initial electron and summed over the polarizations of the two final particles. Similarly, in the same kinematical region, but with the final electron in a bound state, the cross section $d\sigma_{bf}/d\Omega_\gamma$ of bremsstrahlung (radiative recombination), where $d\Omega_\gamma$ is the solid angle corresponding to the photon momentum $\bm{k}$, is given by Eq. (\ref{sigmaeebffinal}). Finally, we mention that similar calculations for the spectrum of photoionization at high photon energies have been performed in Ref. \cite{Pratt_1960} and in Ref. \cite{Pratt_1960b} for the electron being initially in the $K$ and in the $L$ shell, respectively.

\section{Conclusion}

In the present paper we have calculated analytically the cross section $d\sigma/d\bm p_+$ of $e^+e^-$ photoproduction in a Coulomb field exactly in the parameter $\eta=Z\alpha$. The result has been obtained at $\omega\gg m$ and $\epsilon_-\gtrsim m$ (slow electron) and under the assumption $\omega\theta\gg \epsilon_-$. In a wide region of values of $\beta_-=p_-/\epsilon_-$, our results differ essentially from those obtained in the Born approximation. Only in a very narrow region close to the point $\beta_-=1$, the Coulomb corrections vanish. Analogous results concerning the Coulomb corrections are obtained in the complementary case in which the created electron is fast, with the important difference that Coulomb corrections decrease the cross section with respect to the Born value, while they increase it in the case of fast positron.

In the same kinematical region, we have also calculated the cross section $d\sigma_{bf}/d\Omega_+$, when the final electron is in a bound state with arbitrary quantum numbers. The cross section $d\sigma/d\bm k$ of bremsstrahlung in a strong Coulomb field by an ultrarelativistic electron with energy $\epsilon_1$ in the region where the final electron has energy $\epsilon_2\gtrsim m$ coincides with the cross section of $e^+e^-$ photoproduction at $\epsilon _-\gtrsim m$ (slow electron) after the replacement $\beta_-\rightarrow\beta_2$ and $\omega\rightarrow \epsilon_1$. 

Our results are obtained for a pure Coulomb field. However, the effects of screening for high-energy photoproduction in our kinematical region are expected to be important only in the very narrow region close to the point $\beta_-=0$ ($\beta_+=0$) in the case of fast positron (electron).

\section*{Acknowledgments}
A. I. M. gratefully acknowledges the Max-Planck-Institute for Nuclear Physics for warm hospitality and financial support during his visit. The work was supported in part by  the Ministry of Education and Science of the Russian Federation and  the Grant 14.740.11.0082 of Federal special-purpose program “Scientific and scientific-pedagogical personnel of innovative Russia”.

\end{document}